# The APM Galaxy Survey – V. Catalogues of Galaxy Clusters

G. B. Dalton, S. J. Maddox*, W. J. Sutherland & G. Efstathiou
*Department of Physics, University of Oxford, Keble Road, Oxford, OX1 3RH. UK.*

**ABSTRACT**

We describe the construction of catalogues of galaxy clusters from the APM Galaxy survey using an automated algorithm based on Abell-like selection criteria. We investigate the effects of varying several parameters in our selection algorithm, including the magnitude range, and radius from the cluster centre used to estimate the cluster richnesses. We quantify the accuracy of the photometric distance estimates by comparing with measured redshifts, and we investigate the stability and completeness of the resulting catalogues. We find that the angular correlation functions for different cluster catalogues are in good agreement with one another, and are also consistent with the observed amplitude of the spatial correlation function of rich clusters.

## 1 INTRODUCTION

Clusters of galaxies have been used widely as tracers of large-scale structure in the Universe. The catalogues of Abell (1958), Zwicky et al. (1968), and Abell, Corwin & Olowin (1989) were constructed by visual inspection of large numbers of photographic survey plates, and provide estimates of cluster richnesses and distances as well as positions. The process of identifying clusters by eye from survey plates is extremely time consuming and each of these surveys took many years to complete. However, such visual selection is subjective and prone to systematic errors, (see for example, Lucey 1983; Soltan 1988; Sutherland 1988; Dekel et al. 1989; Frenk 1989), and without repeating the inspection of the plates it is very hard to quantify such errors. Shectman (1985) analysed the Lick galaxy counts of Seldner et al. (1977) and produced an objectively selected cluster catalogue by searching for peaks in the projected galaxy density field. This automated approach to cluster selection provides greater control over the resulting cluster sample, and so is less prone to many of the problems associated with visual identification of clusters. However, Shectman's catalogue is based on a visually selected galaxy sample and no photometric information is available for the individual galaxies.

Digitised galaxy surveys over large areas of sky have recently been produced from machine measurements of photographic plates (Maddox et al. 1990a; Heydon-Dumbleton, Collins & MacGillivray 1989). These galaxy surveys are free from subjective selection effects and provide suitable data from which to construct automatically selected cluster catalogues (Dalton et al. 1992; Lumsden et al. 1992). The cluster redshift surveys described by Dalton et al. 1992, Dalton et al. (1994b) (Paper IV of this series) and Dalton et al. (1994a) were based on cluster catalogues selected from the APM Galaxy Survey. In this paper we present a detailed account of the construction of the APM cluster catalogues. These catalogues are selected from the APM Galaxy Survey using an objective algorithm which finds clusters of galaxies that satisfy criteria similar to those of the Abell catalogue. Our aim is to investigate the effect of varying the parameters of cluster selection. Section 2 briefly summarizes some relevant aspects of the APM Galaxy Survey. Our selection algorithm is described in Section 3. In Section 4 we discuss the reliability of the cluster distance estimators. We investigate the properties of various catalogues in Section 5. In Section 6 we measure the angular correlation functions of different catalogues and find that they are insensitive to the exact choice of cluster selection. We also show angular correlation functions for cluster samples limited at different distances and compare the results with those predicted by Limber's (1953) equation. We comment on our final choice of catalogue in Section 7 and sumarise our findings in Section 8.

---

* Present Address: Royal Greenwich Observatory, Madingley Road, Cambridge CB3 0EZ

## 2 GALAXY DATA

Our cluster catalogues are selected from the APM Galaxy Survey, which contains over 2 million galaxies to a magnitude limit of $b_J = 20.5$. The survey covers a contiguous area of 4300 square degrees in the southern sky with Galactic Latitude $b \lesssim -40°$. This region appears to have very little galactic obscuration as evidenced by the maps of both the 21 cm emission from neutral hydrogren (Burstein & Heiles 1978) and the IRAS $100\mu$ emission (Rowan-Robinson et al. 1991).

The construction of the survey has been described in detail by Maddox et al. 1990a,1990b (Papers I and II), and a detailed discussion of the photometric quality of the data has been given by Maddox, Efstathiou & Sutherland (1996) (Paper III). We briefly summarize some relevant points here. The survey was constructed by scanning 185 survey plates from the UK Schmidt telescope with the Automatic Plate Measuring (APM) machine in Cambridge. The scan of each plate covers a $5.8° \times 5.8°$ region of sky, with neighbouring plate centres separated by $5°$ leading to $0.8°$ overlaps along plate boundaries. The data for each plate is stored separately to preserve the multiple measurements in the regions of overlap.

The limiting magnitude of the survey ($b_J = 20.5$) is determined by the completeness of the star-galaxy separation algorithm (Paper I). At this limit we estimate that the galaxy catalogue is 90% complete, and that there is a 7% contamination rate from non-galaxy images. Uniform photometry has been obtained by first correcting the magnitudes on each plate to remove vignetting and desensitization. Each plate zero-point is then adjusted so that the scatter in the galaxy magnitudes in plate overlap regions is minimised (Paper II). The resulting mean *rms* magnitude offset in a plate overlap region is found to be 0.057 magnitudes, and this implies that the rms plate zero-point error is 0.04 magnitudes. The survey was constructed with the specific intention of measuring the angular two-point galaxy correlation function on large scales, and so the photometric accuracy and uniformity have been investigated in great detail (Paper III).

## 3 CLUSTER SELECTION ALGORTIHMS

Our main aim in defining a cluster catalogue is to find peaks above some uniform threshold in the spatial density of galaxies. Since we do not know the distance to each galaxy we must use the galaxy density as seen projected on the sky, and this makes it hard to identify peaks in the spatial distribution. For example, a chance alignment of small groups along the line of sight could produce a high peak in the projected density which does not correspond to a true three dimensional cluster. In general, using the projected distribution to identify spatial peaks in a clustered distribution is likely to introduce projection effects which bias the richness and distance estimates of clusters, and distort the apparent clustering between clusters. We have attempted to design a selection algorithm which minimizes these effects and produces a uniform catalogue that is stable with respect to reasonable changes in the selection parameters.

Our approach is to split the selection process into two stages, as represented by the flow chart in Figure 1. The first stage is to select candidate cluster centres, and the second stage is to estimate a richness parameter and a distance for each candidate. In the second step we estimate richness and distance by counting galaxies within a certain radius from each centre, using analogues of Abell's definitions of richness and characteristic apparent magnitude. Since the richness, distance and position of a cluster are related in a complex way, we use an iterative scheme to find a consistent set of measurements for each cluster. We then apply a richness and distance cut to produce a uniform cluster sample.

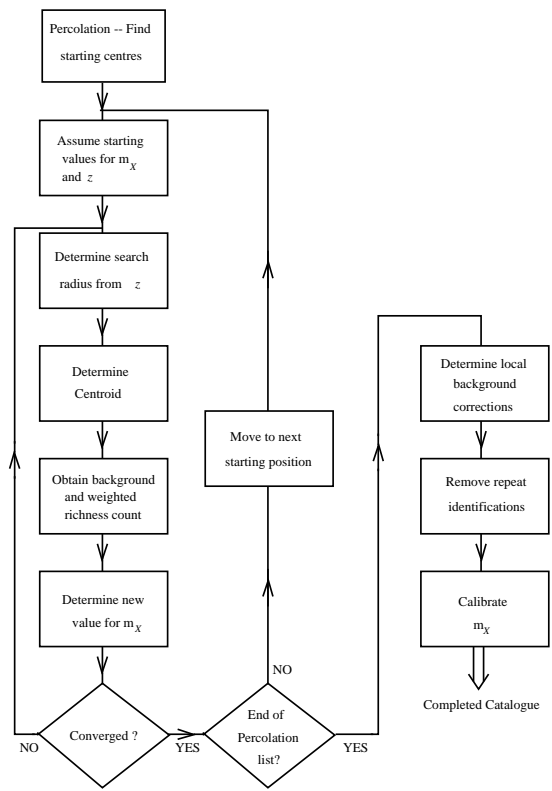

**Figure 1.** Flow diagram representation of the cluster selection procedure (*see text for details*).

### 3.1 Candidate Centres

We tried three different approaches to generate lists of candidate cluster positions.

Our favoured approach is to use a percolation technique in which all pairs of galaxies closer than 0.7 times the mean galaxy separation are linked, and galaxies that are mutually linked are assigned to the same group. The centroid of any group containing $\geq 20$ members is taken as a candidate cluster centre. Another approach that we tried is to construct a smoothed map of the galaxy distribution and then use the positions of peaks in the galaxy density as the candidate cluster list. The third possibility is to start the iterations from a grid of positions chosen so that every point on a plate was contained within the initial counting radius.

We perferred to use the percolation method rather than peak finding because it does not require smoothing of the galaxy data. Using a grid required many more starting positions compared to the percolation groups or peaks and therefore required more time to execute. The extra candidate positions did not lead to significant changes in the final cluster catalogues, and so we chose to use the percolation algorithm in the analysis below.

We apply the percolation algorithm to each plate separately and include a 1.5° border which is drawn from the neighbouring plates. This means that the complete candidate list contains multiple entries for groups in plate overlaps: we remove multiple entries at a later stage of the process. Retaining the overlap area means that clusters close to plate edges are not missed unless they are significantly larger than 1.5°. In general, using tight groups as starting centres will bias our catalogue against clusters which have a large angular extent. Since these clusters are nearby ($z \lesssim 0.035$) they are already well documented in existing catalogues as they are easy to detect visually. These nearby clusters do not have a significant effect on angular or redshift-space correlations since the number of such clusters, typically 8 clusters with $z \leq 0.045$, is a small fraction of any large volume-limited sample.

### 3.2 Counting Radii

Abell's estimates of richness and distance for each cluster depend on counting galaxies within a circle around the cluster. Abell chose the angular radius to be $1.7'/z$ where $z$ is the photometrically estimated cluster redshift. The Abell angular radius corresponds to a physical radius $r_A = 1.5 h^{-1} Mpc$ at the distance of the cluster.

It has been suggested that such a large counting circle in the Abell catalogues leads to significant projection effects caused by overlapping clusters (Sutherland 1988; Dekel et al. 1989; Frenk 1989). Since the apparent galaxy overdensity increases rapidly towards the centre of a cluster, we expect that any projection effects would become much less significant as the counting radius is decreased. With this in mind we have investigated cluster catalogues using smaller counting radii, $r_C$, varying from 0.5 $h^{-1}$Mpc up to Abell's value of 1.5 $h^{-1}$Mpc, which we denote by $r_A$. We have also tested the effect of weighting the galaxy counts as a function of radius from the cluster centre, which can also reduce the effective counting radius.

### 3.3 Richness and Distance Estimators

Having chosen a suitable physical counting radius and candidate centre, we estimate the distance and richness of the cluster. In the Abell catalogues the distance is estimated from the apparent magnitude of the $10^{th}$ brightest cluster galaxy, $m_{10}$, which implicitly requires a self consistent counting radius and distance. The Abell richness is estimated from the number of cluster galaxies within the projected counting radius that are not more than 2 magnitudes fainter than the $3^{rd}$ brightest galaxy.

For our catalogue, we begin by setting the counting circle around each cluster to a projected radius $r_C$ based on an initial guess of $z = 0.1$. The galaxies inside the counting circle are ordered by magnitude and then for each galaxy in the circle we define a cumulative sum:

$$C_i = \frac{1}{\langle w \rangle} \sum_{j \leq i} w_j - \pi r_C^2 \overline{n}(<m_i). \tag{1}$$

The first term is the weighted sum of galaxies brighter than $m_i$ and the second term is the expected background count brighter than $m_i$. The weight function $w$ is set either to $w_i = 1$, or to the function

$$w_i = (1 + 2r_i/r_C)^{-1}, \tag{2}$$

where $r_i$ is the projected distance of the $i$th galaxy from the current cluster centre. The function of equation (2) gives more weight to galaxies near the centre of a cluster, acknowledging the fact that galaxies closer to the centre have a higher probability of cluster membership. In equation 1: $\langle w \rangle$ is the average value of $w$ for a uniform distribution within $r_C$. For the function given by equation (2): $\langle w \rangle = 1 - \frac{1}{2}\ln 3$.

The value of $C_i$ represents the rank of galaxy $i$ in the cluster, i.e. if $C_i = 10$ then the tenth brightest *cluster* galaxy has a magnitude $m_i$. In general we define $m_X$ to be the value of $m_i$ for which $C_i = X$, so that for $X = 10$, $m_X$ is equivalent to Abell's $m_{10}$. We use $m_X$ as our distance estimator, and have investigated several choices of $X$, which we discuss in Section 4. We estimate the cluster richness, $\mathcal{R}$, as the weighted count above background in a range of apparent magnitudes around the characteristic magnitude $m_X$.

For each candidate cluster we iterated from our initial guess by determining the cluster richness, $\mathcal{R}$, computing a new $m_X$, and then determining a new counting radius, $r_C$, based on $m_X$ (see below and Figure 1). A new cluster centre was then defined by the centroid of the galaxy positions within $r_C$. The iterations were stopped if the values of $m_X$ in successive iterations agreed to within 0.025 mag, and if the centres agreed to within $\approx 40''$. Clusters were abandoned if no $m_X$ could be found brighter than the survey limit of $b_J = 20.5$, and flagged if they failed to converge after 20 iterations. Failure was usually due to the cluster oscillating between two positions or two values of $m_X$, or, in the case of very poor clusters, to the cluster position wandering across the plate.

In estimating the richness, $\mathcal{R}$ for each cluster we used either the range $[m_X - 0.5, m_X + 1.5]$, or $[m_X - 0.5, m_X + 1.0]$. If the faint end of the range drops below the magnitude limit of the survey at any point in the iterative process then the cluster is rejected as being either too poor or too distant. Hence the narrower range should enable us to find more distant clusters from the magnitude limited galaxy survey, but at the expense of larger Poisson counting errors in the richness estimate.

Abell and ACO used the galaxy count in the range $[m_3, m_3 + 2]$ as a basis for determining cluster richness counts, and the magnitude of the 10 th brightest cluster galaxy, $m_{10}$, to estimate a distance to the cluster. Scott (1957) noted that one would expect $m_{10}$ to correspond to a brighter absolute magnitude in a rich cluster compared to a poor cluster, since the difference in richness corresponds approximately to a difference in the local normalisation of the luminosity function. Abell's use of a fixed number of galaxies to determine the cluster distances thus gives rise to the "Scott Effect"; the distances to rich clusters are systematically underestimated and the distances to poor clusters are systematically overestimated, with the consequence that the richness estimates are systematically higher for rich clusters and lower for poor clusters due to changes in the projected radii used to determine the richness counts (Sutherland 1989). We have avoided this effect by adjusting the value of $X$ to be a fixed fraction of the cluster richness,

$$X = \max(5, \mathcal{R}/\kappa), \tag{3}$$

where the lower limit on $X$ is imposed to ensure stability of $m_X$ for very poor clusters and $\kappa$ is a constant for a given catalogue. Defined in this way $m_X$ corresponds to a fixed absolute magnitude independant of richness. This absolute magnitude depends on $X$, and can be estimated by adopting a Schechter (1976) form for the luminosity function of cluster galaxies:

$$\phi(L)dL = \phi_*(L/L_*)^{-\alpha} \exp-(L/L_*)dL/L_*, \tag{4}$$

where $L_*$ is a characteristic luminosity, $\alpha$ is the slope at faint luminosities, and $\phi_*$ is the normalisation parameter. The absolute magnitude corresponding to $m_X$ satisfies:

$$\int_{M_X}^{\infty} \phi(M)dM = \frac{1}{\kappa} \int_{\Delta M} \phi(M)dM, \tag{5}$$

where $\phi(M)$ is the luminosity function, expressed in terms of absolute magnitudes.

For each choice of $\kappa$ and magnitude range, we solve equation 5 to determine $M_X$. For $M_X \lesssim M_*$ the distance estimates are prone to large uncertainties due to small number statistics, but for $M_X >> M_*$ the volume surveyed by the catalogue will be significantly restricted. Values of $M_X$ corresponding to various catalogues are given in Table 1 for $\alpha = 1.0$ and 1.2. A Schechter form has been assumed for the galaxy luminosity function. For field galaxies the Schecter function parameters are observed to be (Loveday et al. 1992):

$$M_* = -19.6 \pm 0.13$$
$$\phi_* = (1.4 \pm 0.17) \times 10^{-2} h^3 \mathrm{Mpc}^{-3}$$
$$\alpha = -0.97 \pm 0.15,$$

while for a sample of 14 clusters, Colless (1989) finds:

$$M_* = -20.1 \pm 0.7,$$

for $\alpha$ fixed at -1.25.

### 3.4  Background Corrections

In projection a cluster is seen against the "background" distribution of field galaxies which is clustered and so highly non-uniform. Different estimates of the field galaxy surface density around a cluster can lead to different estimates of the cluster richness. We have estimated the background density around each cluster using either a fixed global estimate of galaxy density based on the observed galaxy number counts as a function of apparent magnitude (Maddox et al. 1990c), or a locally defined value, adjusted for each cluster.

Throughout the iterative part of the selection process (Figure 1) we use the global background estimate. Using a global background correction ensures that the background correction of a cluster is not affected by any other nearby clusters. We obtain a local background estimate for each cluster by determining the surface density within an annulus of 5–6× the counting radius using an equal area projection map of the whole APM survey using $14'$ pixels. This local surface density is used to adjust the estimated richness count to allow for the local background.

**Table 1.** Absolute magnitudes corresponding to $m_X$ for various selection parameters, $\alpha$ is the slope parameter for a Schechter luminosity function.

| $\alpha$ | $X$ | Richness slice | $M_X - M_*$ |
|---|---|---|---|
| 1.0 | $\mathcal{R}/4$ | $[m_X - 0.5, m_X + 1.5]$ | 0.09 |
| | | $[m_X - 0.5, m_X + 1.0]$ | −0.38 |
| | $\mathcal{R}/3$ | $[m_X - 0.5, m_X + 1.5]$ | 0.41 |
| | | $[m_X - 0.5, m_X + 1.0]$ | −0.08 |
| | $\mathcal{R}/2$ | $[m_X - 0.5, m_X + 1.5]$ | 1.00 |
| | | $[m_X - 0.5, m_X + 1.0]$ | 0.45 |
| 1.2 | $\mathcal{R}/4$ | $[m_X - 0.5, m_X + 1.5]$ | 0.34 |
| | | $[m_X - 0.5, m_X + 1.0]$ | −0.22 |
| | $\mathcal{R}/3$ | $[m_X - 0.5, m_X + 1.5]$ | 0.73 |
| | | $[m_X - 0.5, m_X + 1.0]$ | 0.12 |
| | $\mathcal{R}/2$ | $[m_X - 0.5, m_X + 1.5]$ | 1.45 |
| | | $[m_X - 0.5, m_X + 1.0]$ | 0.72 |

In our definition of a local background, we reject cells of the map if they fall within the counting radius of any clusters above a 'global' richness threshold of 10 counts. If we increase the threshold then fewer clusters are excluded from the background estimates; this may lead to the overestimation of the true background level and remove real clusters that are close to other rich clusters. If we decrease the threshold, more structure is excluded from the background estimates, and so they will be more like the global estimate. We checked the effect of this rejection procedure by also generating catalogues with a local background determined from the whole map with no rejected cells.

Table 2 gives the labels assigned to various choices of cluster selection parameters. Each label takes the form of a number to denote the size of the cluster radius, $r_C$, in $h^{-1}$Mpc, followed by three letters denoting the definition of richness, $\kappa$ (C,E...J), the type of background correction used ('G' for Global, 'L' for Local), and the width of the magnitude range used to define the cluster richness ('S' for a 2 magnitude range and 'D' for a 1.5 magnitude range). The choice of cluster selection parameters associated with each catalogue are given in columns (2)–(4). The ALS catalogue listed in the first row was generated with no central weighting; all other catalogues use the central weighting of equation 2 in the determination of both $m_X$ and $\mathcal{R}$.

### 4  DISTANCE CALIBRATIONS

To obtain an estimate of the cluster redshift from $m_X$, we calibrated the $\log z$–$m_X$ relation as follows: An initial catalogue was generated assuming the relation for $m_{10}$ obtained by Postman, Geller & Huchra (1986) for all Abell clusters with measured redshifts. We adapted this relation to $b_J$ magnitudes by assuming a mean $b_J - R$ of 1 mag, and we adopted an initial guess of $m_X \approx m_{10}$. We then obtained a best fit of $\log z$ on $m_X$ by cross-referencing this initial APM catalogue with a list of Abell cluster redshifts within the APM Survey region,

Table 2. The selection parameters used in the construction of the cluster catalogues discussed in the text. Entries are (1) the label used to identify the catalogue; (2) the value of $r_C$ used; (3) The magnitude, $m_f$, of the faint limit of the range used to determine the cluster richness; (4) The value of $\kappa$ used for the richness determination. Columns (5)–(12) list the distance calibration parameters obtained for each catalogue as discussed in section 4: (5) the number of clusters for which redshifts were available for matching within $r_C$; (6) offset to give the absolute magnitude of a standard candle based on a least-squares fit to equation 7; (7) *rms* scatter of the fit; (8) the cut-off redshift. Columns (9)–(12) repeat columns (5)–(8) using $r_C/3$ as a matching radius and a lower magnitude limit of $b_J = 17$ for the fit.

| (1) | (2) | (3) | (4) | (5) | (6) | (7) | (8) | (9) | (10) | (11) | (12) |
|---|---|---|---|---|---|---|---|---|---|---|---|
| Label | $r_C$ | $m_f - m_X$ | $\kappa$ | All ($15.0 \leq m_x \leq 19.0$) | | | | $r_C/3$ ($17.0 \leq m_x \leq 19.0$) | | | |
| | $h^{-1}$ Mpc | | | $N_z$ | $\Delta m$ | $\sigma_{\mathrm{Log}(z)}$ | $z_{lim}$ | $N_z$ | $\Delta m$ | $\sigma_{\mathrm{Log}(z)}$ | $z_{lim}$ |
| 1.5ALS | 1.5 | 1.5 | 4.0 | 71 | -0.945 | 0.140 | 0.100 | 37 | -0.968 | 0.084 | 0.099 |
| 1.5CGS | 1.5 | 1.5 | 4.0 | 73 | -0.883 | 0.118 | 0.102 | 42 | -0.826 | 0.079 | 0.105 |
| 1.5CLS | 1.5 | 1.5 | 4.0 | 73 | -0.852 | 0.125 | 0.104 | 43 | -0.855 | 0.084 | 0.104 |
| 1.0CLS | 1.0 | 1.5 | 4.0 | 76 | -0.602 | 0.114 | 0.114 | 40 | -0.742 | 0.071 | 0.108 |
| 0.75CLS | 0.75 | 1.5 | 4.0 | 82 | -0.508 | 0.120 | 0.119 | 48 | -0.688 | 0.069 | 0.111 |
| 0.5CLS | 0.5 | 1.5 | 4.0 | 84 | -0.299 | 0.152 | 0.129 | 55 | -0.523 | 0.087 | 0.118 |
| 1.5CLD | 1.5 | 1.0 | 4.0 | 52 | 0.063 | 0.092 | 0.177 | 22 | -0.208 | 0.047 | 0.160 |
| 1.0CLD | 1.0 | 1.0 | 4.0 | 65 | 0.174 | 0.101 | 0.184 | 33 | -0.055 | 0.082 | 0.170 |
| 0.75CLD | 0.75 | 1.0 | 4.0 | 78 | 0.184 | 0.126 | 0.185 | 32 | -0.086 | 0.073 | 0.168 |
| 0.75ELD | 0.75 | 1.0 | 2.0 | 88 | -0.869 | 0.137 | 0.125 | 56 | -1.120 | 0.086 | 0.114 |
| 0.75FLD | 0.75 | 1.0 | 3.0 | 87 | -0.094 | 0.116 | 0.167 | 46 | -0.414 | 0.080 | 0.149 |
| 0.75GLD | 0.75 | 1.0 | 2.5 | 94 | -0.389 | 0.125 | 0.150 | 59 | -0.680 | 0.085 | 0.135 |
| 0.75HLD | 0.75 | 1.0 | 2.25 | 94 | -0.578 | 0.137 | 0.140 | 65 | -0.839 | 0.080 | 0.127 |
| 0.75ILD | 0.75 | 1.0 | 2.2 | 87 | -0.633 | 0.125 | 0.137 | 53 | -0.879 | 0.085 | 0.125 |
| 0.75JLD | 0.75 | 1.0 | 2.1 | 89 | -0.652 | 0.145 | 0.136 | 58 | -0.926 | 0.080 | 0.122 |
| 0.5CLD | 0.5 | 1.0 | 4.0 | 93 | 0.289 | 0.108 | 0.192 | 56 | -0.052 | 0.080 | 0.170 |

$$z_{est} = 10^{\alpha m_X - \delta}; \quad \alpha = 0.2, \delta = 4.707, \qquad (6)$$

which corresponds to an estimated redshift of $\approx 0.125$ for $m_X = 19$. The redshift list was obtained from the Andernach (1991) compilation, excluding all clusters referenced as Fairall & Jones (1988), or those from ACO without an entry in the ACO catalogue 'previous' column, since the redshifts for these clusters were obtained by cross referencing with galaxy redshift lists and so may be contaminated by foreground galaxies. We exclude all clusters with discordant redshift entries, since these may well be confused superpositions and would contaminate the calibration. We also removed any cluster for which the documented position lies in one of the APM Survey holes. For the catalogues discussed here we then included further redshifts for a sample of 32 clusters in the APM Survey region measured with AUTOFIB at the Anglo Australian Telescope in October 1989, and 7 additional clusters from Muriel, Nicotra & Lambas 1990,1991. We analysed all these clusters using a maximum likelihood fit to the apparent magnitude distribution (Dalton et al. 1994b) to remove any remaining contamination due to poorly determined cluster redshifts. The final list contains 107 clusters with reliable redshift determinations.

For a standard candle, the relationship between absolute magnitude, $M$, and apparent magnitude, $m$, as a function of redshift is given by

$$m(z) = M_* - \Delta m + 25 + 5\log(d_l(z)) + K_z z, \qquad (7)$$

where $K_z$ is the reddening correction and $d_l(z)$ is the luminosity distance defined by

$$d_l(z) = \frac{2c}{H_0}(1+z)[1-(1+z)^{-1/2}],$$

assuming a critical density, spatially flat Universe. $\Delta m$ is the apparent magnitude difference between the object to be used as a standard candle and the apparent magnitude corresponding to $M_*$ at the relevant redshift. We cross-referenced the list of available redshifts with each catalogue in turn, and obtained a calibration of the $m_X$ distance indicator by fitting equation 7 using $M_* = -19.6$ and $K_z = 3$ (Efstathiou, Ellis & Peterson 1988) with $\Delta m$ as a free parameter. The catalogue depth is then the calibrated redshift at which the faint end of the magnitude slice used to determine $\mathcal{R}$ equals the magnitude limit of the APM Survey ($b_J = 20.5$).

The calibration parameters for the various catalogues are given in columns (5)–(12) of Table 2.

Column (7) of Table 2 lists the *rms* scatter in the fit for each catalogue. We investigated the effect of contamination of the magnitude–redshift relation by cluster misidentifications by reducing the matching radius used to identify clusters from the literature with a cluster center from $r_C$ to $r_C/3$. We find that this procedure greatly reduces the scatter of the fits, without significantly changing the depth estimates. The effect of this on the 1.5CLS catalogue is shown in Figure 2. The large scatter introduced by matching to existing data within a full cluster radius is present over the whole range of $m_X$ shown.

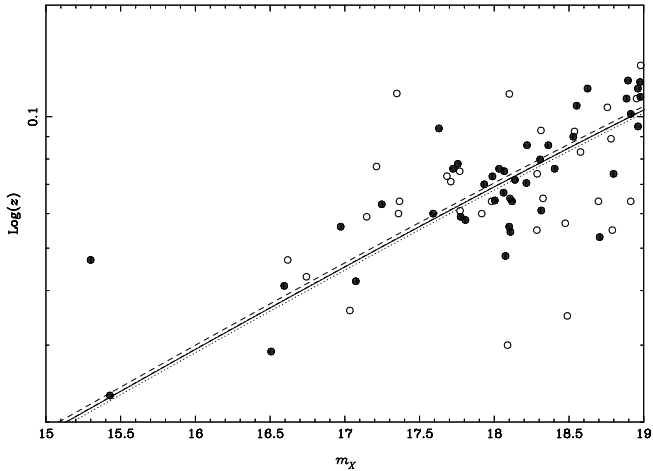

**Figure 2.** The $m_X$ vs $\log z$ relation for clusters in the 1.5CLS catalogue. The symbols refer to clusters matched to existing redshift data for $r_{match} = r_C/3$ (●), and additional clusters matched for $r_{match} = r_C$ (○). The lines show least-squares fits of the theoretical relation (equation 7) to all points (dot-dashed line), and to ● only (dashed line). The dotted and solid lines show the same fits if we reject points brighter than $m_X = 17$.

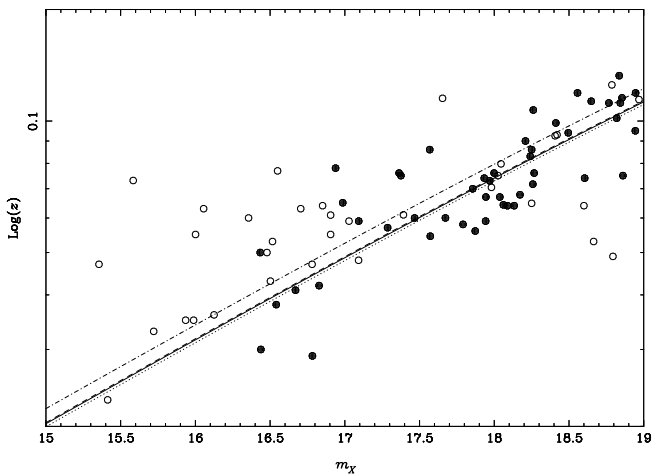

**Figure 3.** The $m_X$ vs $\log z$ relation for clusters in the 0.75CLS catalogue. The symbols and line styles correspond to those in Figure 2. The scatter at bright $m_X$ is clearly smaller if we consider only those clusters matched within $r_C/3$.

As we move to catalogues with smaller search radii (Figure 3) then we see that the scatter is still present, but that it is now dominated by clusters with bright $m_X$. Reducing the matching radius to $r_C/3$ now has a much smaller effect at faint $m_X$, as would be expected, since we are already using a smaller matching radius by means of our small defining radius for the cluster. The increased scatter at bright $m_X$ results from large fluctuations in the number of background objects found in a narrow magnitude range within a small radius. As this is likely to affect only a small number of nearby clusters it should not be a major problem for studies of large-scale structure based on our catalogue, but should be taken into account for determinations of the distance calibration.

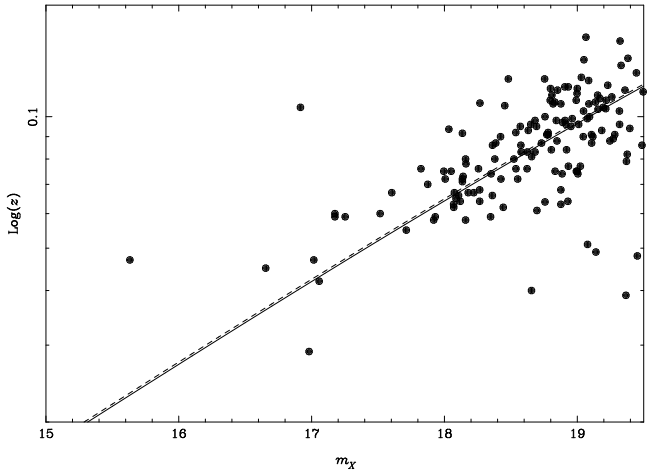

**Figure 4.** The $m_X$ vs $\log z$ relation for clusters in the parent catalogue used for the redshift survey of Dalton et al. (1994a). The solid line shows the fit to the whole range of $m_X$ while the dashed line shows the fit for $m_X \geq 17$

We therefore restricted the fits to $m_X \geq 17$ (columns 9–12 of Table 2).

From column (12) of Table 2 we see the effect of the central weighting procedure on equation 5. The values of $\Delta m$ given are expected to differ from the numbers in column (4) of Table 1, since the value of $M_*$ appropriate to a galaxy cluster is somewhat brighter than the value determined for field galaxies (Colless 1989). The difference in the calibrated value of $\Delta m$ obtained by applying the central weighting procedure is only 0.1 mag. The values of $M_*$ inferred from the fits show a trend to fainter $M_*$ as we reduce the cluster search radius, but the differences in $\Delta m$ obtained between different values for $X$ are consistent with the predictions of equation 5 given in Table 1.

As a conclusion to this part of our study, we note that a search radius of 0.75 $h^{-1}$Mpc seems preferable to other choices, since at 1.0 $h^{-1}$Mpc there is still some residual scatter in the magnitude–redshift relation characteristic of a large radius (as in Figure 2), but at 0.5 $h^{-1}$Mpc the fluctuations in the richness counts of nearby clusters become more pronounced, affecting an increasingly large fraction of the catalogue.

A further question associated with the redshift calibrations is that of the useful depth of the catalogue. The catalogues discussed so far have been limited to $m_X \leq 19.0$ by the extent of the magnitude slice used to determine the cluster richness. We therefore generated a similar set of catalogues using a magnitude slice of $[m_X - 0.5, m_X + 1.0]$ from which to determine our cluster richness estimates. This allows us to include clusters with $m_X \leq 19.5$. However, from the fits to the $r_C = 0.75$ $h^{-1}$Mpc catalogues listed in table 2, we conclude that changing the magnitude slice does indeed give us an increase in the effective depth of the catalogues. However, from Table 1 we now expect $m_X$ to be brighter than $M_*$ if we keep $X = \mathcal{R}/4$. Increasing $X$ to account for the reduced size of the richness slice brings $m_X$ back to just fainter than $M_*$ as desired, but reduces the effective depth of the catalogue.

The completion of the first phase of our redshift survey (Paper IV) provides additional data to check the above cal-ibrations. In Figure 4 we show the fit of equation 7 to those clusters in the 0.75JLD catalogue with $\mathcal{R} \geq 50$ matched to an updated list of redshifts. The parameters of the fit to the data in the range $17.0 \leq m_X \leq 19.5$ are

$$\Delta m = -1.03; \sigma_{\mathrm{Log}(z)} = 0.101; z_{lim} = 0.117.$$

The group of five clusters located at the lower right of Figure 4 represent some of the low redshift clusters that are misidentified by our algorithm, as discussed in Section 3.

## 5 ERROR ESTIMATES AND COMPLETENESS

We search each catalogue for repeat cluster identifications. These arise when the algorithm converges to the same region of the plate from different starting points. We use these data to obtain an estimate of the intrinsic errors in our determinations of the cluster $m_X$, $\mathcal{R}$, and the position of the cluster centre. We note that these errors reflect only the uncertainties intrinsic to the cluster selection algorithm, and that the true uncertainties will be slightly larger due to the photometric uncertainties in the galaxy data (Paper III). Intrinsic errors for some of our catalogues are shown in columns 5,7, and 8 of Table 3. In each case where a cluster is detected more than once we retain only the richest occurence.

Table 3 also lists the mean values of $\mathcal{R}$ and $m_X$ obtained for subsamples of these catalogues with the same estimated depth and space density. We define subsamples for a group of catalogues by determining the limiting value of $m_X$, $m_{lim}$ for each catalogue which corresponds to the same calibrated depth as the full magnitude limit of the shallowest catalogue in the group. To fix the same estimated space density for each subsample in the group we then sort each catalogue on richness and select the richest 500 clusters brighter than $m_{lim}$. The richness limits obtained in this way are listed in column 3 of table 3. These numbers imply that the estimates of $\mathcal{R}$ and $m_X$ are uncertain at the 10% and 0.25 mag levels while the typical positional accuracy should be better than $3'$ for $r_C = 0.75$ $h^{-1}$ Mpc. The intrinsic scatter in the richness estimates implies that the membership of a subsample will depend on the particular group of catalogues under consideration.

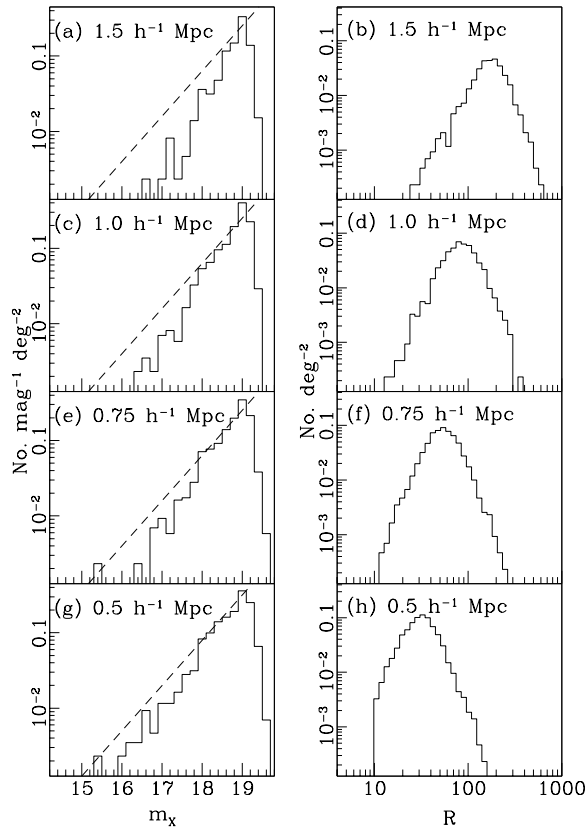

**Figure 5.** The differential number counts of clusters as a function of $m_X$ and $\mathcal{R}$ for the CLS catalogues listed in rows 1–4 of Table 3. For the number-magnitude counts we have chosen a richness limit in each catalogue which fixes the estimated space density to that of the 500 richest clusters brighter than $m_X = 19.0$ in the shallowest catalogue (see text). The dashed lines represent a slope of 0.6 with arbitrary normalisation. For the number-richness relations we have limited each catalogue to $m_X = 19.0$

**Table 3.** Error estimates for $m_X$ and $\mathcal{R}$ based on cluster multiple detections. Entries in columns (3),(4) and (6) are given for the richest 500 clusters in each catalogue brighter than $m_{lim}$, where $m_{lim}$ is chosen to give the same calibrated depth in each catlogue. Entries in columns (5),(7) and (8) give the error estimates in $m_X$, $\mathcal{R}$, and the cluster position based on clusters found with multiple entries. The samples listed in rows 1–4 are matched to the 1.5CLS sample, and the others are matched to the 0.75ELD sample.

| (1) Sample | (2) $m_{lim}$ | (3) $\mathcal{R}_{min}$ | (4) $\langle m_X \rangle$ | (5) $\sigma_{m_X}$ | (6) $\langle \mathcal{R} \rangle$ | (7) $\sigma_{\mathcal{R}}$ | (8) $\sigma_P(°)$ |
|---|---|---|---|---|---|---|---|
| 1.5CLS   | 19.0  | 190  | 18.6  | 0.32 | 247.2 | 28.89 | 0.157 |
| 1.0CLS   | 18.89 | 99   | 18.4  | 0.28 | 134.5 | 11.75 | 0.085 |
| 0.75CLS  | 18.83 | 67   | 18.3  | 0.28 | 89.4  | 6.55  | 0.053 |
| 0.5CLS   | 18.66 | 39   | 18.0  | 0.25 | 60.4  | 4.43  | 0.031 |
| 0.75CLD  | 18.48 | 34.6 | 17.68 | 0.29 | 55.7  | 6.39  | 0.052 |
| 0.75FLD  | 18.81 | 39.0 | 18.00 | 0.24 | 66.9  | 5.70  | 0.049 |
| 0.75GLD  | 19.07 | 43.2 | 18.34 | 0.30 | 71.8  | 6.04  | 0.043 |
| 0.75HLD  | 19.22 | 44.7 | 18.53 | 0.27 | 74.9  | 5.60  | 0.044 |
| 0.75ILD  | 19.26 | 45.6 | 18.56 | 0.27 | 74.5  | 5.62  | 0.043 |
| 0.75JLD  | 19.31 | 46.0 | 18.61 | 0.28 | 76.3  | 5.07  | 0.040 |
| 0.75ELD  | 19.50 | 51.2 | 18.78 | 0.29 | 79.1  | 5.18  | 0.040 |

We have investigated the completeness of each catalogue by considering the cluster number counts as functions of $m_X$ and $\mathcal{R}$. Figure 5 shows the differential number counts for the CLS catalogues listed in the first four rows of table 3. For a homogeneous sample of clusters we would expect the number-magnitude relation to have a slope of 0.6 provided that cosmological effects and evolution can be neglected. Figures 5(c), (e), and (g) follow the predicted relation closely at faint magnitudes, but with evidence for incompleteness at low redshift which increases with $r_C$.

The number-richness relations shown in figure 5 are all well represented by power laws of slope $-4.5$ up to the point at which the distributions rapidly become incomplete. These figures show that the limit of 500 clusters that we have chosen for comparison purposes remains above the completeness limit for each catalogue, but that we detect more clusters as we reduce the finding radius, consistent with our interpretation of $m_X$ as a distance estimator.

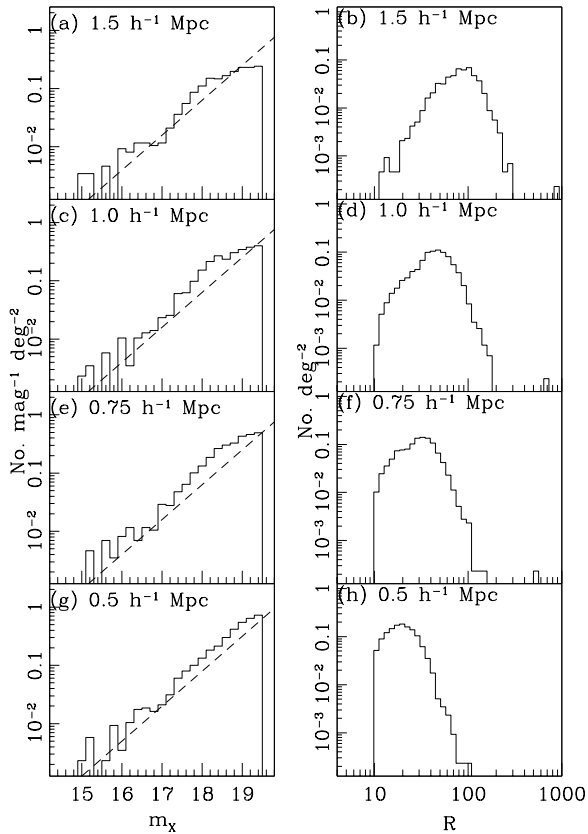

**Figure 6.** The number counts for the CLD catalogues corresponding to those shown in figure 5 and limited to the same effective depth. The richness limits for the matched subsamples shown here are 68, 39.5, 29.3, and 18.6, respectively. The number-richness histograms are for limiting magnitudes of $m_X = 19.5$.

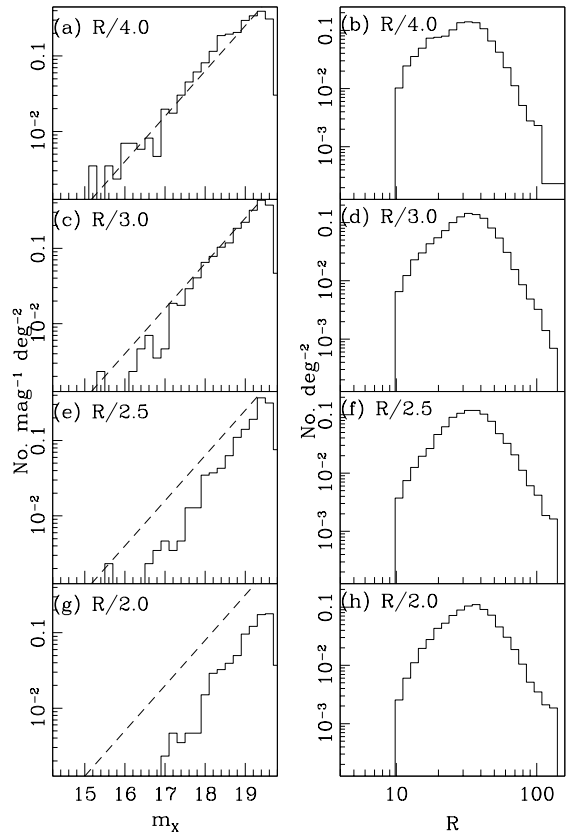

**Figure 7.** Number counts for a subset of the 0.75LD samples listed in table 3. As figure 5. The $N(\mathcal{R})$ histograms are for all clusters brighter than $m_X = 19.5$.

Figure 6 shows similar distribution functions for the deeper CLD catalogues. The subsamples shown here have been limited to the same depth as those shown in figure 5. Comparing the richness limits given in the figure caption with the number-richness relations shows that these samples are reaching the completeness limits for the catalogues. The number-magnitude histograms show little evidence of incompleteness at bright magnitudes, which is surprising given that we expect incompleteness at low-redshift to result from the percolation stage of the selection process. This illustrates the problem of choosing $X$ such that $m_X$ is too close to the bright end of the luminosity function. As well as introducing large uncertainties into $M_X$ itself, and so reducing the effectiveness of the distance calibration, this has the effect of pushing the richness counting range to very bright magnitudes for nearby clusters and so introduces large statistical uncertainties into the background correction for a significant fraction of the catalogue.

Figure 7 shows the effect of varying $\kappa$. The variations in the number-magnitude histograms reflect the different depths of the catalogues, but the additional depth gained by reducing the width of the richness counting slice increases the effect of the incompleteness at low redshifts. The richness completeness limits are similar for all catalogues with the same detection radius.

## 6 THE ANGULAR CORRELATION FUNCTION

As our motivation for selecting uniform cluster catalogue is to provide a tool for the study of large-scale structure it is important to ensure that our method of selecting clusters does not affect the clustering properties of the final catalogue. We therefore estimate the angular two-point correlation function for each catalogue using the estimator:

$$w_{cc}(\theta) = 2f\frac{(DD)}{(DR)} - 1, \qquad (8)$$

where $f$ is the ratio of the number of random points to the number of clusters in the sample. The random catalogue consists of 20 000 points distributed uniformly within the survey area. Errors are estimated using the formula $\delta w_{cc} = (1+w_{cc})/\sqrt{(DD)}$, but the error bars plotted in figures 8–13 are $2\delta w_{cc}$ to compensate for the fact that this formula tends to underestimate the true errors for strongly clustered data.

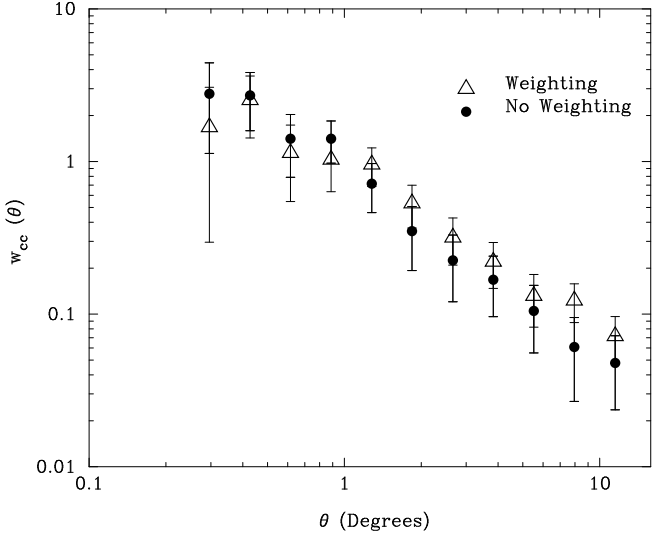

**Figure 8.** The sensitivity of the angular clustering to the weighting scheme: A comparison of the angular correlation functions for similar catalogues generated with (1.5ALS) and without (1.5CLS) the weight function of equation 2 applied to the galaxy counts (filled and open symbols, respectively).

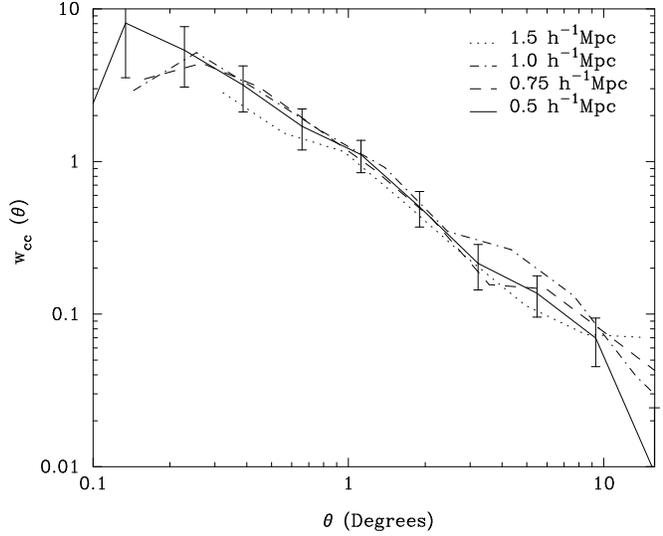

**Figure 10.** The sensitivity of the angular clustering to the cluster detection radius: The angular correlation functions for cluster samples generated using $X = \mathcal{R}/4$ and a wide (2 mag) range for the richness determination. For clarity the error bars are shown only for the sample with $r_C = 0.5\ h^{-1}$ Mpc (solid line).

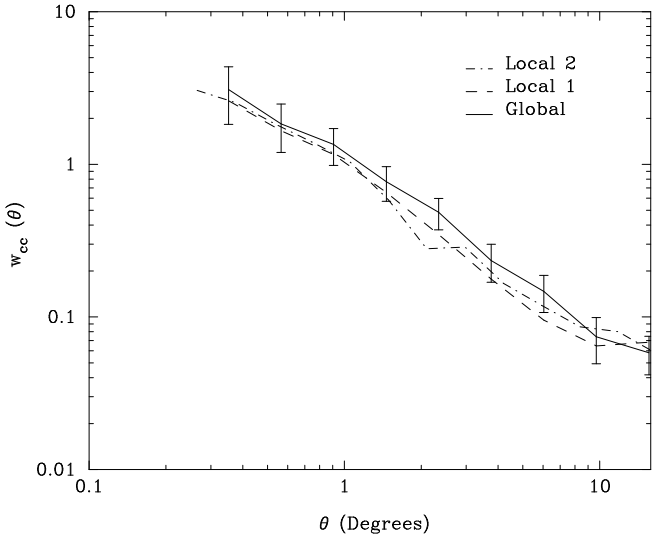

**Figure 9.** The sensitivity of the angular clustering to the background correction: A comparison of the angular correlation functions for similar catalogues generated using global (1.5CGS, solid line) and local (1.5CLS, dashed line) background corrections. The dot-dashed line shows the results obtained by including other clusters in the local background estimate, as discussed in the text.

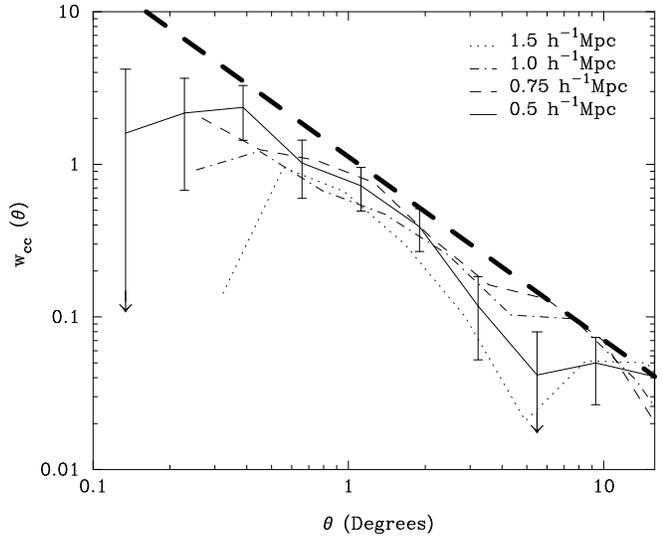

**Figure 11.** Angular correlation functions for cluster samples similar to those of Figure 10 but generated using a narrow (1.5 mag) range for the richness determination. The limiting depths of these samples have been chosen to match those of Figure 10. The heavy dashed line shows $w_{cc}(\theta) = (\theta/1.1°)^{-1.2}$, which provides a good fit to the data shown in Figure 10.

### 6.1 Comparisons between catalogues

The angular correlation functions are expected to scale with the depth of the samples in accordance with Limber's equation (e.g. Peebles 1980, equation 51.7). In comparing the results from different catalogues it is therefore necessary to take account of the limiting depth, which can be computed from the magnitude–redshift relation as discussed in Section 5.

Figure 8 shows the effect of the galaxy weighting scheme. The amplitude of $w_{cc}$ on scales $\gtrsim 1°$ appears to be slightly larger when weighting is used, although the estimates of $w_{cc}$ for the two weighting schemes are consistent with one another to within the error bars on all scales shown.

In Figure 9 we show the angular correlation functions measured for catalogues constructed with local and global background corrections. On large scales ($\theta \gtrsim 2°$) the global richness counts give estimates of $w_{cc}$ that are enhanced by a factor of $\sim 1.4$. A comparison of the dashed and dot-dashed lines in Figure 9 shows that the procedure of rejecting cluster centres from the local background determination has no detectable effect on the large scale clustering properties.

Figure 10 shows the variation in the estimates of $w_{cc}$ obtained by changing the cluster radius used in the selection. There is no systematic trend with $r_C$ for these samples, suggesting that the use of a large radius to define clusters does not seriously bias the clustering properties. In Figure 11 we show results for similar catalogues generated using a narrower magnitude range to define the cluster richness. For comparison purposes we used subsamples generated to the same effective depth as those in Figure 10. The most striking feature here is that all these catalogues give slightly lower correlation amplitudes than those of Figure 10. The explanation for this behaviour follows from the discussion of Section 4: For $X = \mathcal{R}/4$ the values of $\Delta m$ in equation 7 are close to zero when a narrow magnitude range is used for the richness, and so $m_X$ is unlikely to be a useful distance estimator for these catalogues. If the distance estimates for each iteration of the selection process are poorly defined then the algorithm becomes sensitive to the presence of groups of galaxies in projection which are not associated along the line of sight. The presence of such groups in the catalogue dilutes the observed clustering. This effect becomes more significant for large $r_C$ due to the larger projected area covered by each cluster candidate.

The effect of varying $\kappa$ is shown in figure 12 for catalogues with $r_C = 0.75\,h^{-1}$Mpc and the narrow magnitude range for the cluster richness definition. There is no indication here that the choice of $\kappa$ affects the clustering properties. In comparing the data for $\kappa = \mathcal{R}/4$ (filled circles) to that shown by the dashed line in Figure 11 it should be noted that the subsample considered here is effectively deeper ($m_X \geq 18.47$) than that shown in Figure 11 ($m_X \geq 18.23$). A subsample from $\kappa = \mathcal{R}/2.1$ catalogue from this group, but taken to the same effective depth as the samples shown in Figures 10 and 11 gives a correlation function that lies on top of the data shown in Figure 10. This suggests that the tendency for lower correlation amplitudes for the $CLD$ samples relative to the $CLS$ samples, shown in Figure 11, is more significant at brighter magnitudes and low richnesses.

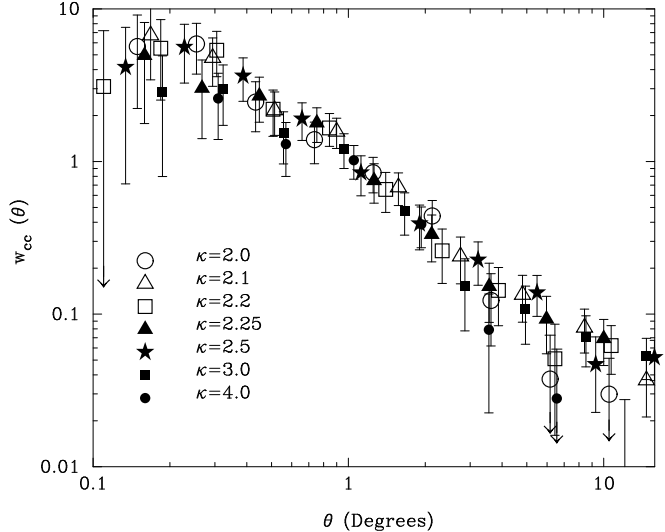

**Figure 12.** The angular correlation functions for subsamples of clusters with $r_C = 0.75\,h^{-1}$ Mpc and a narrow (1.5 mag) range for the cluster richness.

### 6.2 Scaling Properties

As a further test of the quality of our final catalogue, we investigate the scaling properties of the angular correlation functions for two different samples of clusters selected from our final choice of catalogue. We use the penultimate catalogue from Table 3 as this forms the basis for the redshift survey described by Dalton et al. (1994a). Figure 7$l$ shows that this catalogue should be complete for $\mathcal{R} \geq 40$. Combining this with the magnitude completeness limit of $m_X \leq 19.5$ we obtain a sample of 1205 clusters. In figure 13 we show the results for the 400 brightest and 400 faintest clusters from this sample.

For each subsample we estimate the redshift distribution by applying equation 7 and broadening the distribution with the observed scatter in the fit listed in Table 2. Figure 14 shows that the redshift distributions predicted in this way are in good agreement with the true distributions obtained from redshift samples over the same range of $m_X$. The predicted distribution is shown for clusters with $\mathcal{R} \geq 40$ and $m_X \leq 19.2$ to match the range of $m_X$ covered by the redshift survey, but to a lower richness limit. The predicted distributions obtained in this way are then smoothed with a Gaussian of width 4000km s$^{-1}$. We use these smoothed distributions in the relativistic version of Limber's equation:

$$w(\theta) = \frac{\int_0^\infty \left(\frac{dN}{dz}\right)^2 \left(\frac{dz}{dx}\right) dx \int_{-\infty}^\infty du\,\xi(r,t)}{\left(\int_0^\infty \left(\frac{dN}{dz}\right) dz\right)^2}, \qquad (9)$$

where $x$ is co-ordinate distance. The $u$ integral in equation 9 is evaluated over a model for the spatial correlation function with the physical separation of two objects given by

$$r = (1+z)^{-1}(u^2/F^2 + x^2\theta^2)^{1/2}, \quad u = x_2 - x_1, \qquad (10)$$

where the function $F(x)$ relates $x$ to the proper radial distance, (Peebles, equation 56.16),

$$F(x) = [1 - y^2(\Omega_0 - 1)]^{1/2}, \quad y = H_0 x/c. \qquad (11)$$

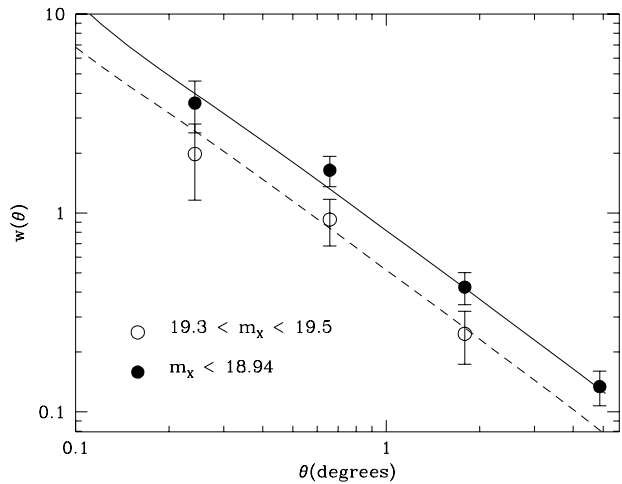

**Figure 13.** The angular correlation functions for two subsamples drawn from the catalogue used in the redshift survey of Dalton et al. (1994a) (see text). The solid and dotted lines show the predictions for the power-law fit to spatial correlation function observed by Dalton et al. (1994a).

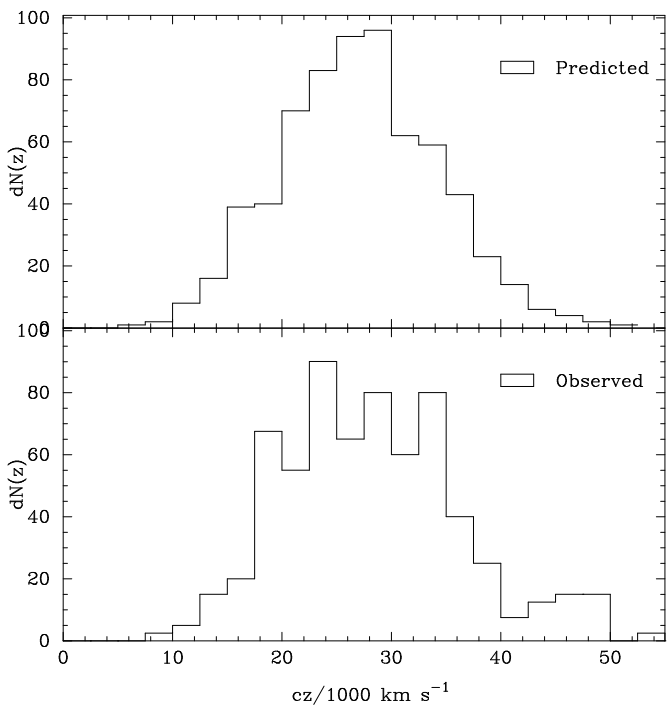

**Figure 14.** The redshift distribution predicted from the magnitude–distance relation of the parent catalogue (see text) compared to the redshift distribution observed by Dalton et al. (1994a). The observed distribution has been normalised to the same number of clusters as the predicted distribution.

$\frac{dN}{dz}$ is the expected number–redshift relation for the sample used to determine $w$. We assume $\Omega_0 = 1$ and we approximate $\xi_{cc}$ by the form

$$\xi_{cc}(r,z) = B r^{-\gamma}(1+z)^{-\gamma}, \tag{12}$$

where the factor of $(1+z)^{-\gamma}$ corresponds to constant $\xi$ at fixed comoving separation.

The solid and dotted lines in figure 13 show the predictions for the two estimates of $w_{cc}$ using the estimated redshift distributions and the best fit power law model for $\xi_{cc}$ obtained for this catalogue by Dalton et al. (1994a):

$$\xi_{cc}(s) = \left(\frac{s}{r_0}\right)^{-\gamma}, \quad \gamma = 2.14, \quad r_0 = 14.3\ h^{-1}\,\mathrm{Mpc}.$$

The angular correlation functions lie within the range of values allowed by the observational limits on $\xi_{cc}$. These results can be compared with those of Sutherland & Efstathiou (1991), who investigated the scaling properties of clusters drawn from different distance classes of the Abell catalogue and found the amplitudes of the angular correlation functions to be a factor of $\gtrsim 2$ larger than those predicted by the observed spatial correlation function. Sutherland & Efstathiou (1991) concluded that this observation supported the evidence for inhomogeneities in Abell's selection procedure giving rise to spurious clustering. Our findings here give support to the conclusions of Dalton et al. (1994a) that no such effects are present at significant levels in the APM cluster catalogues.

## 7 CHOICE OF THE FINAL CATALOGUE

From the analyses of Sections 4 and 6 we conclude that the final choice of catalogue should have little effect on the final clustering properties of a redshift sample, other than the limiting depth. For the purposes of the APM Cluster Redshift Survey we chose to adopt a small cluster radius, $r_C = 0.75\ h^{-1}\,\mathrm{Mpc}$ to reduce the sensitivity of the selection algorithm to spurious line-of-sight associations, and the combination of a narrow (1.5 mag) range to determine our cluster richnesses with a small value of $\kappa$ (2.1) so as to retain the low-redshift stability of the catalogue. The complete 0.75JLD catalogue of 957 clusters is given in Appendix A.

## 8 DISCUSSION

We have described a procedure for selecting clusters of galaxies from the APM Galaxy Survey. The procedure is based on Abell's original classification procedure, but with modifications which are designed to improve the accuracy and remove biases in the photometrically derived distances of rich clusters. We have investigated the effect of varying the input parameters of the selection algorithm, and find that the effective depth of the catalogue may be increased significantly, but that this gain is at the expense of reducing the sensitivity of the algorithm to nearby clusters.

We derive a distance calibration for each catalogue using redshifts for Abell clusters drawn from the literature. We find that usingm only those redshifts from the literature that match within a fraction of a cluster radius greatly reduces the scatter in the distance calibration at bright magnitudes.

We have investigated the completeness limits of the various possible catalogues and find that the use of a large cluster radius reduces the sensitivity to poor clusters and can introduce serious contamination from line of sight associations. A cluster radius as small as $r_C = 0.5\ h^{-1}$Mpc can give rise to large uncertainties in the distance calibration which are amplified by the iterative selection algorithm. For our final catalogue we therefore adopt $r_C = 0.75\ h^{-1}$Mpc with a 1.5 magnitude range about $m_X$ to define the richness to maximise the depth of our survey. From the data in Table 2, and from Figure 12 it is apparent that the results are fairly insensitive to changes in $\kappa$ in the range $2.0 \lesssim \kappa \lesssim 2.5$. All the catalogues show a slight incompleteness for very nearby clusters, which is caused by the percolation stage of the selection algorithm and arises because nearby clusters have a large angular extent and a low surface density contrast on the sky.

We have evaluated the angular correlation functions for a number of catalogues, and find that the clustering properties are insensitive to the way in which clusters are selected, provided that we retrict our catalogues to a reasonable range of input parameters. We find that the scaling properties of the angular correlation functions for the catalogue used by Dalton et al. (1994a) as the basis of their redshift survey are in good agreement with the predictions obtained from Limber's equation.

The additional observation that the angular clustering seen in the APM cluster catalogues appears to be stable to changes in the cluster selection method suggests that the underlying cause of the observed inhomogeneities in the Abell catalogues may be due to systematic variations in the quality of the galaxy catalogues used in their construction (Dalton 1992).

## 9  ACKNOWLEDGEMENTS

GBD acknowledges reciept of a SERC studentship. WJS acknowledges reciept of a PPARC advanced fellowship. GPE acknowledges reciept of a PPARC senior fellowship.

## APPENDIX A1: THE APM CLUSTER CATALOGUE

In this appendix we present a listing of clusters in the catalogue used as the basis for the redshift survey of Dalton et al. (1994a). On the basis of Figure 7 we restrict this list to 957 clusters with $\mathcal{R} \geq 40$ and $m_X \leq 19.4$. The catalogue is given in Table A1: Column (1) gives an identifier for each cluster. Columns (2) and (3) list the cluster R.A. and Dec. in 1950 coordinates in the form hhmmss.s and ddmmss. Column (5) lists the projected radius that was used in the final iteration of the cluster finding algorithm. Column (6) gives $z_{est}$ as determined from $m_X$ using equation 7, while Column (7) gives the cluster richness. Notes in column (7) refer to visual inspection of the clusters on the UKSTU survey plates:

a This cluster is close to a bright star which has been excised from the galaxy catalogue. A part of the cluster area is therefore missing from the survey data and so the richness may have been underestimated.

b The field of this cluster includes a $2'$ foreground galaxy $0.7 r_C$ from the centre which has been broken up into several small galaxian images by the APM software. These additional objects may have caused the richness to be slightly overestimated.

c This field is dominated by a $2'$ foreground galaxy which has been broken up into several small galaxian images by the APM software. The field does not appear to contain a true cluster. The cluster was below the richness threshold used by Dalton et al. (1994a) and was not visually inspected at the time of those observations.

Notes to column (8) denote the source of the redshift:

1 Redshift is from Paper IV.

2 Redshift measured by Dalton et al. (1994a). Note that this survey was limited to $m_X \leq 19.2$.

3 Redshift measured as part of the deep extension to the APM cluster redshift survey (Croft et al. 1996 *in preparation*).

4 Cluster is found in the Edinburgh–Milano Cluster Redshift Survey (Collins et al. 1995). If only one reference is present then the redshift is adopted from this source, otherwise this entry indicates that the cluster is also found in that survey.

5 cluster redshift has been drawn from the Las Campanas Redshift Survey (Shectman et al. 1996, LCRS).

† For the cluster APM100 we have adopted the value given by Collins et al. (1995) as they point out this is in better agreement with the maximum likelihood redshift estimate quoted in Paper IV than our single galaxy redshift.

We find 73 clusters in this list in common with Collins et al. (1995). For 51 of these we have independent redshift determinations using the method described by Dalton et al. (1994b). In only one case (APM100) do we find a discrepancy between our redshift determinations and the multi-object observations of Collins et al. (1995).

We find the 56 of our clusters can be unambiguously identified from the LCRS, by which we mean that three or more galaxies are found with concordant redshifts within the projected radius of the cluster. 27 of these represent new cluster redshifts.

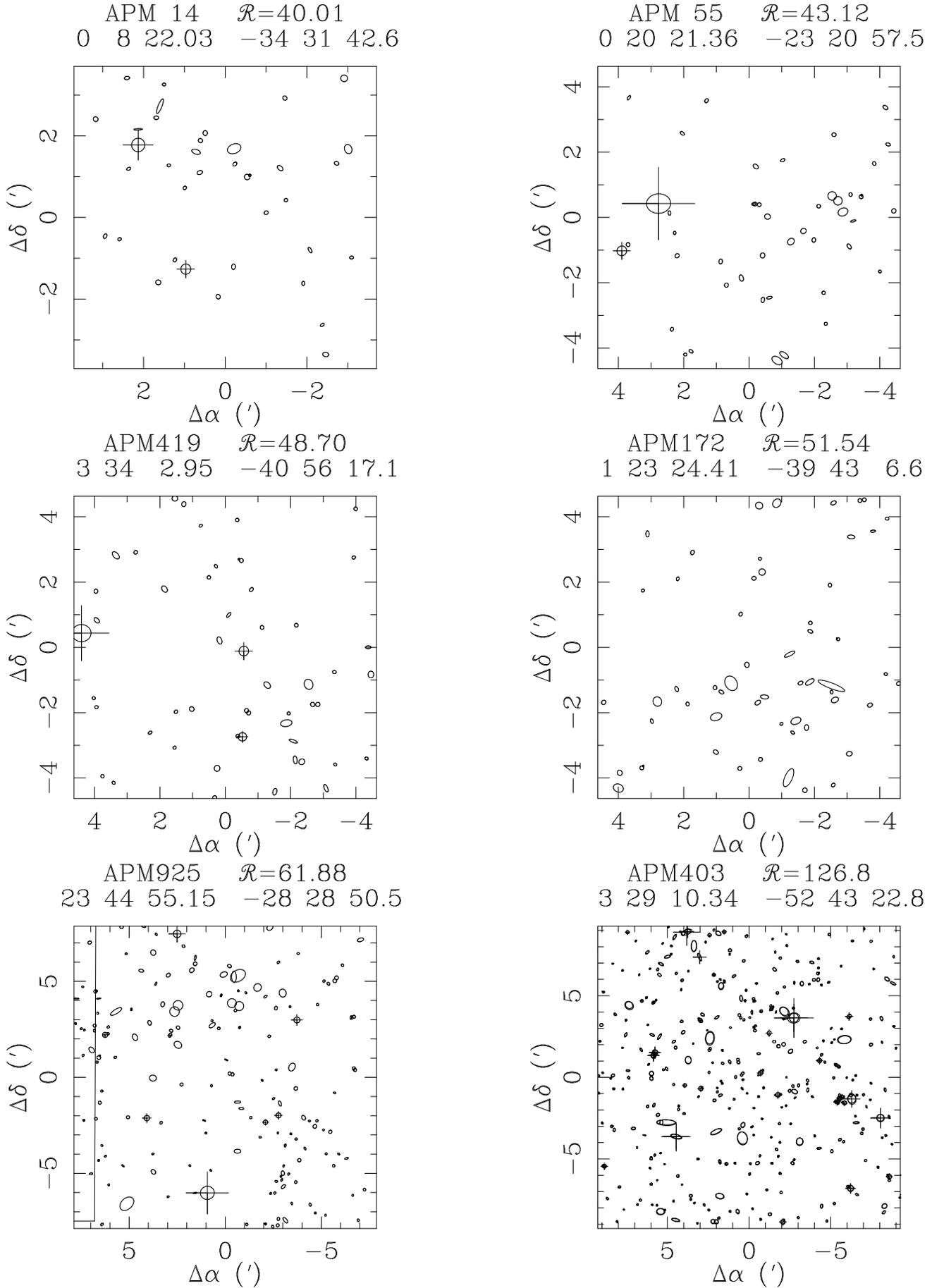

**Figure A1.** A selection of clusters taken at random from the catalogue, in order of increasing richness. The ellipses represent image sizes and shapes as detected by the APM analysis software. Objects classified as stars are denoted by crosses, but are limited to those stars with $b_{\rm J} \lesssim 17$. The box at the left of the lower left panel denotes the boundary of part of the survey mask.